\documentstyle[12pt,epsf]{article}

\renewcommand{\bar}[1]{\overline{#1}}
\newcommand{\M}{{\cal M}}
\newcommand{\VEV}[1]{\left\langle{#1}\right\rangle}
\newcommand{\etal}{{\em et al.}}
\newcommand{\ie}{{\it i.e.}}
\newcommand{\eg}{{\it e.g.}}

\newcommand{\ket}[1]{\vert\,{#1}\rangle}

\begin{document}
\begin{flushright}
SLAC--PUB--8235\\
July 1999
\end{flushright}
\bigskip\bigskip

\thispagestyle{empty}
\flushbottom

\begin{center}
{{\Large \bf Hard Exclusive and Diffractive Processes in QCD}\footnote{\baselineskip=14pt
     Work supported by the Department of Energy, contract
     DE--AC03--76SF00515.}}\\
\vspace{1.0cm}
Stanley J. Brodsky\\
{\it{Stanford Linear Accelerator Center\\
Stanford University,
Stanford, California 94309}}\\
e-mail: sjbth@slac.stanford.edu\\
\end{center}
\vfill
\begin{center}
Abstract
\end{center}
Exclusive and semi-exclusive processes,  the diffractive dissociation of hadrons
into jets, and hard diffractive processes such as vector meson leptoproduction
provide new testing grounds for QCD and essential information on the
structure of light-cone wavefunctions of hadrons, particularly the pion
distribution amplitude.  I review the basic features of the leading-twist QCD
predictions and the problems and challenges of studying QCD at the amplitude
level.  The application of the light-cone formalism to the exclusive
semi-leptonic
decay of heavy hadrons is also discussed.

\vfill
\begin{center}
Presented at a workshop on \\
Exclusive and Semiexclusive Processes at High Momentum Transfer\\
Jefferson Laboratory \\
20-22 May 1999
\end{center}
\vfill

\newpage

\section{Introduction}

Exclusive hard-scattering reactions and hard diffractive reactions are now
providing
an invaluable window into the structure and
dynamics of hadronic amplitudes.  Recent measurements of the
photon-to-pion transition form factor at CLEO,\cite{Gronberg:1998fj} the
diffractive dissociation of pions into jets at Fermilab,\cite{E791}
diffractive vector meson leptoproduction at Fermilab and HERA, and the new
program
of experiments on exclusive proton and deuteron processes at Jefferson
Laboratory
are now yielding fundamental information on hadronic wavefunctions,
particularly the
distribution amplitude of mesons.  Such information is also critical for
interpreting exclusive heavy hadron decays and the matrix elements and
amplitudes
entering $CP$-violating processes at the $B$ factories.

The natural formalism for
describing the hadronic wavefunctions which enter exclusive and
diffractive amplitudes is the light-cone
Fock representation obtained by quantizing the theory at fixed
at fixed ``light-cone" time
$\tau = t+z/c$.\cite{PinskyPauli} This representation is the
extension of Schr\"odinger many-body theory to the relativistic domain.
In a quantum
theory a bound state cannot have a fixed number of constituents.  For
example, the
proton state has the Fock expansion
\begin{eqnarray}
\ket p &=& \sum_n \VEV{n\,|\,p}\, \ket n \nonumber \\
&=& \psi^{(\Lambda)}_{3q/p} (x_i,\vec k_{\perp i},\lambda_i)\,
\ket{uud} \\[1ex]
&&+ \psi^{(\Lambda)}_{3qg/p}(x_i,\vec k_{\perp i},\lambda_i)\,
\ket{uudg} + \cdots \nonumber
\label{eq:b}
\end{eqnarray}
representing the expansion of the exact QCD eigenstate on a non-interacting
quark and gluon basis.  The probability amplitude
for each such
$n$-particle state of on-mass shell quarks and gluons in a hadron is given by a
light-cone Fock state wavefunction
$\psi_{n/H}(x_i,\vec k_{\perp i},\lambda_i)$, where the constituents have
longitudinal light-cone momentum fractions
$
x_i ={k^+_i}/{p^+} = (k^0_i+k^z_i)/(p^0+p^z)\ , \sum^n_{i=1} x_i= 1
$,
relative transverse momentum
$\vec k_{\perp i} \ , \sum^n_{i=1}\vec k_{\perp i} = \vec 0_\perp$,
and helicities $\lambda_i.$ The effective lifetime of each configuration
in the laboratory frame is ${2 P_{lab}/({\M}_n^2- M_p^2}) $ where
$
\M^2_n = \sum^n_{i=1}(k^2_{\perp i} + m^2_i)/x_i < \Lambda^2 $
is the off-shell invariant mass and $\Lambda$ is a global
ultraviolet regulator.  A crucial feature of the light-cone formalism is
the fact that the form of the
$\psi^{(\Lambda)}_{n/H}(x_i,
\vec k_{\perp i},\lambda_i)$ is invariant under longitudinal boosts; \ie,\ the
light-cone wavefunctions expressed in the relative coordinates $x_i$ and
$k_{\perp i}$ are independent of the total momentum
$P^+$,
$\vec P_\perp$ of the hadron.
The ensemble
\{$\psi_{n/H}$\} of such light-cone Fock
wavefunctions is a key concept for hadronic physics, providing a conceptual
basis for representing physical hadrons (and also nuclei) in terms of their
fundamental quark and gluon degrees of freedom.  Given the
$\psi^{(\Lambda)}_{n/H},$ we can construct any spacelike electromagnetic or
electroweak form factor from the diagonal overlap of the LC
wavefunctions.\cite{BD}
Similarly, the matrix elements of the currents that define quark and gluon
structure functions can be computed from the integrated squares of the LC
wavefunctions.\cite{LB}

There has been much progress analyzing
exclusive and diffractive reactions at large momentum transfer from first
principles
in QCD.  Rigorous statements can be made on the basis of asymptotic freedom and
factorization theorems which separate the underlying hard quark and gluon
subprocess amplitude from the nonperturbative physics incorporated into the
process-independent hadron distribution amplitudes
$\phi_H(x_i,Q)$,\cite{LB} the valence light-cone
wavefunctions integrated over $k^2_\perp<Q^2$.  An important new
application is the
recent analysis of hard exclusive
$B$ decays by Beneke, {\it et al.}\cite{Beneke:1999br} Key features of such
analyses are: (a) evolution equations for distribution amplitudes which
incorporate the operator product expansion, renormalization group
invariance, and conformal symmetry;
\cite{LB,Brodsky:1980ny,Brodsky:1986ve,Muller:1994hg,Ball:1998ff,Braun:1999te}
(b) hadron helicity conservation which follows from the underlying chiral
structure of QCD;\cite{Brodsky:1981kj} (c) color transparency, which
eliminates corrections to hard exclusive amplitudes from initial and final state
interactions at leading power and reflects the underlying gauge theoretic basis
for the strong interactions;\cite{BM,Frankfurt:1992dx} and (d) hidden color
degrees of freedom in nuclear wavefunctions, which reflects the color
structure of hadron and nuclear wavefunctions.\cite{bjl83} There have also been
recent advances eliminating renormalization scale ambiguities in hard-scattering
amplitudes via commensurate scale
relations\cite{Brodsky:1995eh,Brodsky:1996tb,Brodsky:1999gm} which connect the
couplings entering exclusive amplitudes to the
$\alpha_V$ coupling which controls the QCD heavy quark
potential.\cite{Brodsky:1998dh} The postulate that the QCD coupling has an
infrared fixed-point provides an understanding of the applicability of
conformal scaling and dimensional counting
rules to physical QCD processes.\cite{BF,Matveev:1973ra,Brodsky:1998dh} The
field of analyzable exclusive processes has recently been expanded to a new
range
of QCD processes, such as electroweak decay
amplitudes, highly virtual diffractive processes such as
$\gamma^* p \to \rho p$,\cite{Brodsky:1994kf,Collins:1997hv} and semi-exclusive
processes such as
$\gamma^* p \to \pi^+ X$ \cite{acw,Brodsky:1998sr,BB} where the $\pi^+$ is
produced in isolation at large $p_T$.

\begin{figure}[htb]
\begin{center}
\leavevmode
\epsfbox{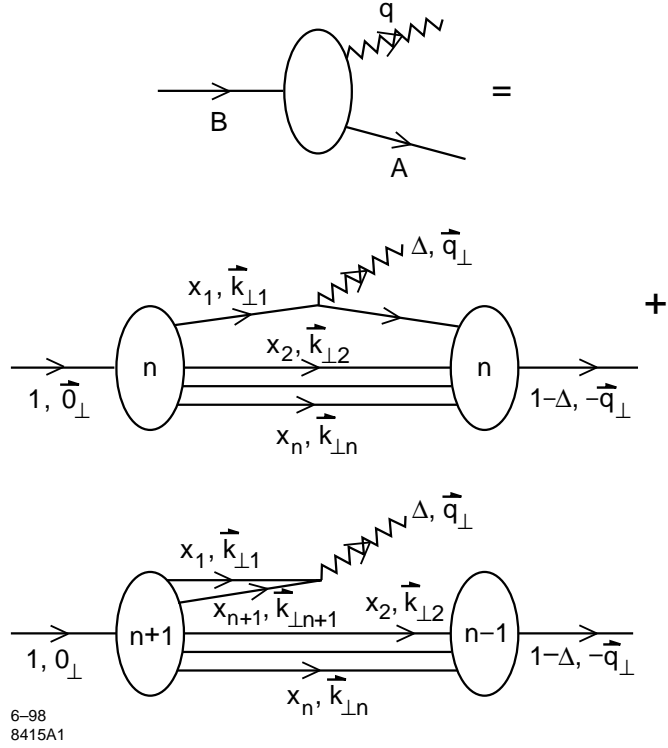}
\end{center}
\caption[*]{Exact representation of electroweak decays and time-like form
factors in the
light-cone Fock representation.
}
\label{fig1}
\end{figure}

\section{Electoweak Decays and the Light-Cone Fock Expansion}

Dae Sung Hwang and I have recently shown how
exclusive semi-leptonic $B$-decay amplitudes, such as
$B\rightarrow A \ell \bar{\nu}$ can be evaluated exactly in the light-cone
formalism.\cite{BrodskyHwang} These timelike decay matrix elements require the
computation of
the diagonal matrix element $n \rightarrow n$ where parton number is conserved,
and the off-diagonal $n+1\rightarrow n-1$ convolution where the current
operator
annihilates a $q{\bar{q'}}$ pair in the initial $B$
wavefunction.  See Fig. \ref{fig1}.  This term is a consequence of the fact
that the
time-like decay $q^2 = (p_\ell + p_{\bar{\nu}} )^2 > 0$
requires a positive light-cone momentum fraction
$q^+ > 0$.  Conversely for space-like currents, one can choose $q^+=0$, as in
the Drell-Yan-West representation of the space-like electromagnetic form
factors.\cite{DY,BD,West} However, the off-diagonal convolution can yield
a nonzero
$q^+/q^+$ limiting form as $q^+ \rightarrow 0$.  This extra term appears
specifically
in the case of ``bad" currents such as $J^-$ in which the coupling to $q\bar q$
fluctuations in the light-cone wavefunctions are favored.  In effect, the $q^+
\rightarrow 0$ limit generates
$\delta(x)$ contributions as residues of the $n+1\rightarrow n-1$
contributions.
The necessity for this zero mode $\delta(x)$ terms has been noted by
Chang, Root and Yan,\cite{CRY} and Burkardt.\cite{BUR}

The off-diagonal $n+1 \rightarrow n-1$ contributions provide a new
perspective for the physics of $B$-decays.  A semi-leptonic decay involves
not only matrix
elements where a quark changes flavor, but also a contribution where the
leptonic pair is
created from the annihilation of a $q {\bar{q'}}$ pair within the Fock
states of the initial $B$
wavefunction.  The semi-leptonic decay thus can occur from the annihilation of a
nonvalence quark-antiquark pair in the initial hadron.  This feature will carry
over to exclusive hadronic $B$-decays, such as $B^0 \rightarrow
\pi^-D^+$.  In this case the pion can be produced from the coalescence of a
$d\bar u$ pair emerging from the initial higher particle number Fock
wavefunction of the $B$.  The $D$ meson is then formed from the remaining quarks
after the internal exchange of a $W$ boson.

In principle, a precise evaluation of the hadronic matrix elements needed
for $B$-decays
and other exclusive electroweak decay amplitudes requires knowledge of all of
the light-cone Fock wavefunctions of the initial and final state hadrons.
In the case of model gauge theories such as QCD(1+1) \cite{Horn} or collinear
QCD \cite{AD} in one-space and one-time dimensions, the complete evaluation of
the light-cone wavefunction is possible for each baryon or meson bound-state
using the DLCQ method.\cite{DLCQ,AD} It would be interesting to use such
solutions
as a model for physical $B$-decays.

\section{The Transition from Soft to Hard Physics}

The existence of an exact formalism
provides a basis for systematic approximations and a control over neglected
terms.  For example, one can analyze exclusive semi-leptonic
$B$-decays which involve hard internal momentum transfer using a
perturbative QCD formalism\cite{BHS,Beneke:1999br} patterned after the analysis
of form
factors at large momentum transfer.\cite{LB} The hard-scattering analysis
proceeds
by writing each hadronic wavefunction as a sum of soft and hard contributions
\begin{equation}
\psi_n = \psi^{{\rm soft}}_n (\M^2_n < \Lambda^2) + \psi^{{\rm hard}}_n
(\M^2_n >\Lambda^2) ,
\end{equation}
where $\M^2_n $ is the invariant mass of the partons in the $n$-particle
Fock state and
$\Lambda$ is the separation scale.
The high internal momentum contributions to the wavefunction $\psi^{{\rm
hard}}_n $ can be calculated systematically from QCD perturbation theory
by iterating the gluon exchange kernel.  The contributions from high
momentum transfer exchange to the
$B$-decay amplitude can then be written as a convolution of a hard-scattering
quark-gluon scattering amplitude $T_H$ with the distribution
amplitudes $\phi(x_i,\Lambda)$, the valence wavefunctions obtained by
integrating the
constituent momenta up to the separation scale
${\cal M}_n < \Lambda < Q$.  This is the basis for the
perturbative hard-scattering analyses.\cite{BHS,Sz,BABR,Beneke:1999br}
In the exact analysis, one can
identify the hard PQCD contribution as well as the soft contribution from
the convolution of the light-cone wavefunctions.
Furthermore, the hard-scattering contribution can be systematically improved.

\section{Hard Exclusive Processes}

In general, hard exclusive hadronic amplitudes such as quarkonium
decay, heavy hadron decay,  and scattering amplitudes where hadrons
are scattered with large momentum transfer can be factorized at leading
power as a
convolution of distribution amplitudes and hard-scattering quark/gluon matrix
elements\cite{LB}
\begin{eqnarray}
\M_{\rm Hadron} &=& \prod_H \sum_n \int
\prod^{n}_{i=1} d^2k_\perp \prod^{n}_{i=1}dx\, \delta
\left(1-\sum^n_{i=1}x_i\right)\, \delta
\left(\sum^n_{i=1} \vec k_{\perp i}\right) \nonumber \\[2ex]
&& \times \psi^{(\Lambda)}_{n/H} (x_i,\vec k_{\perp i},\Lambda_i)\,
T_H^{(\Lambda)} \ .
\label{eq:e}
\end{eqnarray}
Here $T_H^{(\Lambda)}$ is the underlying quark-gluon
subprocess scattering amplitude in which the (incident and final) hadrons are
replaced by their respective quarks and gluons with momenta $x_ip^+$, $x_i\vec
p_{\perp}+\vec k_{\perp i}$ and invariant mass above the
separation scale $\M^2_n > \Lambda^2$.
The essential part of the wavefunction is the hadronic distribution amplitudes,
\cite{LB} defined as the integral over transverse momenta of the valence (lowest
particle number) Fock wavefunction; \eg\ for the pion
\begin{equation}
\phi_\pi (x_i,Q) \equiv \int d^2k_\perp\, \psi^{(Q)}_{q\bar q/\pi}
(x_i, \vec k_{\perp i},\lambda)
\label{eq:f}
\end{equation}
where the global cutoff $\Lambda$ is identified with the
resolution $Q$.  The distribution amplitude controls leading-twist exclusive
amplitudes at high momentum transfer, and it can be related to the
gauge-invariant Bethe-Salpeter wavefunction at equal light-cone time
$\tau = x^+$.  The $\log Q$ evolution of the hadron distribution amplitudes
$\phi_H (x_i,Q)$ can be derived from the
perturbatively-computable tail of the valence light-cone wavefunction in the
high transverse momentum regime.  The LC ultraviolet
regulators provide a factorization scheme for elastic and inelastic
scattering, separating the hard dynamical contributions with invariant mass
squared $\M^2 > \Lambda^2_{\rm global}$ from the soft physics with
$\M^2 \le \Lambda^2_{\rm global}$ which is incorporated in the
nonperturbative LC wavefunctions.  The DGLAP evolution of quark and gluon
distributions can also be derived in an analogous way by computing the
variation of
the Fock expansion with respect to $\Lambda^2$.  The natural
renormalization scheme
for the QCD coupling in hard exclusive processes is $\alpha_V(Q)$, the
effective charge defined from the scattering of two infinitely-heavy
quark test charges.  The renormalization scale can then be determined
from the virtuality of the exchanged momentum of the gluons, as in the BLM and
commensurate scale
methods.\cite{BLM,Brodsky:1995eh,Brodsky:1996tb,Brodsky:1999gm}

The features of exclusive processes to leading power in the transferred
momenta are well known:

(1) The leading power fall-off is given by dimensional counting rules for
the hard-scattering amplitude: $T_H \sim 1/Q^{n-1}$, where $n$ is the total
number
of fields
(quarks, leptons, or gauge fields) participating in the hard
scattering.\cite{BF,Matveev:1973ra} Thus the reaction is dominated by
subprocesses
and Fock states involving the minimum number of interacting fields.  The
hadronic
amplitude follows this fall-off modulo logarithmic corrections from the
running of
the QCD coupling, and the evolution of the hadron distribution amplitudes.
In some
cases, such as large angle $p p \to p p $ scattering, pinch contributions from
multiple hard-scattering processes must also be
included.\cite{Landshoff:1974ew}
The general success of dimensional counting rules implies that the
effective coupling
$\alpha_V(Q^*)$ controlling the gluon exchange propagators in
$T_H$ are frozen in the infrared, \ie, have an infrared fixed point, since the
effective momentum transfers $Q^*$ exchanged by the gluons are often a
small fraction
of the overall momentum transfer.\cite{Brodsky:1998dh} The pinch contributions
are suppressed by a factor decreasing faster than a fixed power.\cite{BF}

(2) The leading power dependence is given by hard-scattering amplitudes $T_H$
which conserve quark helicity.\cite{Brodsky:1981kj,Chernyak:1999cj} Since the
convolution of $T_H$ with the light-cone wavefunctions projects out states with
$L_z=0$, the leading hadron amplitudes conserve hadron helicity; \ie, the
sum of
initial and final hadron helicities are conserved.

(3) Since the convolution of the hard scattering amplitude $T_H$ with the
light-cone
wavefunctions projects out the valence states with small impact parameter,
the essential part of the hadron wavefunction entering a hard exclusive
amplitude has
a small color dipole moment.  This leads to the absence of initial or final
state
interactions among the scattering hadrons as well as the color transparency.
of quasi-elastic interactions in a nuclear target.\cite{BM,Frankfurt:1992dx}
For example, the amplitude for diffractive vector meson photoproduction
$\gamma^*(Q^2) p \to \rho p$, can be written as convolution of the virtual
photon and
the vector meson Fock state light-cone wavefunctions the $g p \to g p$
near-forward matrix element.\cite{Brodsky:1994kf} One can easily show that only
small transverse size $b_\perp \sim 1/Q$ of the vector meson distribution
amplitude is involved.  The sum over the interactions of the exchanged
gluons tend to
cancel reflecting its small color dipole moment.  Since the hadronic
interactions are
minimal,  the
$\gamma^*(Q^2) N \to
\rho N$ reaction at large $Q^2$ can occur coherently throughout a nuclear
target in
reactions without absorption or final state interactions.  The $\gamma^*A
\to V A$ process thus provides a natural framework for testing QCD color
transparency.  Evidence for color transparency in such reactions has been
found by Fermilab experiment E665.\cite{Adams:1997bh}

\section{Measurement of Light-cone Wavefunctions and Tests of Color
Transparency via Diffractive Dissociation.}

Diffractive multi-jet production in heavy
nuclei provides a novel way to measure the shape of the LC Fock
state wavefunctions and test color transparency.  For example, consider the
reaction
\cite{Bertsch,MillerFrankfurtStrikman,Frankfurt:1999tq}
$\pi A \rightarrow {\rm Jet}_1 + {\rm Jet}_2 + A^\prime$
at high energy where the nucleus $A^\prime$ is left intact in its ground
state.  The transverse momenta of the jets have to balance so that
$
\vec k_{\perp i} + \vec k_{\perp 2} = \vec q_\perp < {R^{-1}}_A \ ,
$
and the light-cone longitudinal momentum fractions have to add to
$x_1+x_2 \sim 1$ so that $\Delta p_L < R^{-1}_A$.  The process can
then occur coherently in the nucleus.  Because of color transparency,  \ie,
the cancelation of color interactions in a small-size color-singlet
hadron,  the valence wavefunction of the pion with small impact
separation will penetrate the nucleus with minimal interactions,
diffracting into jet pairs.\cite{Bertsch}
The $x_1=x$, $x_2=1-x$ dependence of
the di-jet distributions will thus reflect the shape of the pion distribution
amplitude; the $\vec k_{\perp 1}- \vec k_{\perp 2}$
relative transverse momenta of the jets also gives key information on
 the underlying shape of the valence pion
wavefunction.\cite{MillerFrankfurtStrikman,Frankfurt:1999tq} The QCD
analysis can be
confirmed by the observation that the diffractive nuclear amplitude
extrapolated to
$t = 0$ is linear in nuclear number $A$, as predicted by QCD color
transparency.  The integrated diffractive rate should scale as $A^2/R^2_A \sim
A^{4/3}$.  A diffractive dissociation experiment of this type, E791,  is now in
progress at Fermilab using 500 GeV incident pions on nuclear
targets.\cite{E791} The preliminary results from E791 appear to be consistent
with color transparency.  The momentum fraction distribution of the jets is
consistent with a valence light-cone wavefunction of the pion consistent with
the shape of the asymptotic distribution amplitude, $\phi^{\rm asympt}_\pi (x) =
\sqrt 3 f_\pi x(1-x)$.  As discussed below, data from
CLEO\cite{Gronberg:1998fj} for the
$\gamma
\gamma^* \rightarrow \pi^0$ transition form factor also favor a form for
the pion distribution amplitude close to the asymptotic solution\cite{LB}
to the perturbative QCD evolution
equation.\cite{Kroll,Rad,Brodsky:1998dh,Feldmann:1999wr,Schmedding:1999ap}
It will also be interesting to study diffractive tri-jet production using proton
beams
$ p A \rightarrow {\rm Jet}_1 + {\rm Jet}_2 + {\rm Jet}_3 + A^\prime $ to
determine the fundamental shape of the 3-quark structure of the valence
light-cone wavefunction of the nucleon at small transverse
separation.\cite{MillerFrankfurtStrikman} One interesting possibility is 
that the distribution amplitude of the
$\Delta(1232)$ for $J_z = 1/2, 3/2$ is close to the asymptotic form $x_1
x_2 x_3$,  but that the proton distribution amplitude is more complex.
This would explain why the $p \to\Delta$ transition form factor appears to
fall faster at large $Q^2$ than the elastic $p \to p$ and the other $p \to
N^*$ transition form factors.\cite{Stoler:1999nj}
Conversely, one can use incident real and virtual photons:
$ \gamma^* A \rightarrow {\rm Jet}_1 + {\rm Jet}_2 + A^\prime $ to
confirm the shape of the calculable light-cone wavefunction for
transversely-polarized and longitudinally-polarized virtual photons.  Such
experiments will open up a direct window on the amplitude
structure of hadrons at short distances.

\section {Leading Power Dominance in Exclusive QCD Processes}

There are a large number of measured exclusive reactions in which the
empirical power
law fall-off predicted by dimensional counting and PQCD appears to be
accurate over a large range of momentum transfer.
These include processes such as the proton form factor, time-like meson pair
production in $e^+ e^-$ and $\gamma
\gamma$ annihilation, large-angle scattering processes such as pion
photoproduction
$\gamma p \to \pi^+ p$, and nuclear processes such as the deuteron form
factor at
large momentum transfer and deuteron photodisintegration.\cite{Brodsky:1976rz} A
spectacular example is the recent measurements at CESR of the photon to pion
transition form factor in the reaction $e \gamma \to e
\pi^0$.\cite{Gronberg:1998fj}
As predicted by leading twist QCD\cite{LB} $Q^2 F_{\gamma
\pi^0}(Q^2)$ is essentially constant for 1 GeV$^2 < Q^2 < 10$ GeV$^2.$
Further, the
normalization is consistent with QCD at NLO if one assumes that the pion
distribution
amplitude takes on the form $\phi^{\rm asympt}_\pi (x) =
\sqrt 3 f_\pi x(1-x)$ which is the asymptotic solution\cite{LB} to the
evolution equation for the pion
distribution amplitude.\cite{Kroll,Rad,Brodsky:1998dh,Schmedding:1999ap}

The measured deuteron form factor and the deuteron photodisintegration
cross section
appear to follow the leading-twist QCD predictions at large momentum
transfers in the
few GeV region.\cite{Holt:1990ze,Bochna:1998ca} The normalization of the
measured deuteron form factor is large compared to model calculations
\cite{Farrar:1991qi} assuming that the deuteron's six-quark wavefunction can be
represented at short distances with the color structure of two color
singlet baryons.
This provides indirect evidence for the presence of hidden color components as
required by PQCD.\cite{bjl83}

There are, however, experimental exceptions to the general success of the
leading twist PQCD approach, such as (a) the dominance of the $J/\psi \to
\rho \pi$
decay which is forbidden by hadron helicity conservation and (b) the strong
normal-normal spin asymmetry $A_{NN}$ observed in polarized elastic $p p
\to p p$
scattering and an apparent breakdown of color transparency at large CM
angles and
$E_{CM} \sim 5$ GeV.  These conflicts with leading-twist PQCD predictions can be
used to identify the presence of new physical effects.  For example,
It is usually assumed that a heavy quarkonium state such as the
$J/\psi$ always decays to light hadrons via the annihilation of its heavy quark
constituents to gluons.  However, as Karliner and I \cite{Brodsky:1997fj}
have recently
shown, the transition $J/\psi \to \rho
\pi$ can also occur by the rearrangement of the $c \bar c$ from the $J/\psi$
into the $\ket{ q \bar q c \bar c}$ intrinsic charm Fock state of the $\rho$ or
$\pi$.  On the other hand, the overlap rearrangement integral in the
decay $\psi^\prime \to \rho \pi$ will be suppressed since the intrinsic
charm Fock state radial wavefunction of the light hadrons will evidently
not have nodes in its radial wavefunction.  This observation provides a natural
explanation of the long-standing puzzle why the $J/\psi$ decays prominently to
two-body pseudoscalar-vector final states, whereas the $\psi^\prime$
does not.  The unusual effects seen in elastic proton-proton scattering at
$E_{CM}
\sim 5$ GeV and large angles could be related to the charm threshold
and the effect of a $\ket{ uud uud c \bar c }$ resonance which would appear
as in
the $J=L=S=1$ $p p $ partial wave.\cite{Brodsky:1988xw}

If the pion distribution amplitude is close to its asymptotic form, then one can
predict the normalization of exclusive amplitudes such as the spacelike
pion form factor $Q^2 F_\pi(Q^2)$.  Next-to-leading order
predictions are now becoming available which incorporate higher order
corrections
to the pion distribution amplitude as well as the hard scattering
amplitude.\cite{Muller:1994hg,Melic:1999hg,Szczepaniak:1998sa} However, the
normalization of the PQCD prediction for the pion form factor depends
directly on the
value of the effective coupling
$\alpha_V(Q^*)$ at momenta $Q^{*2} \simeq Q^2/20$.  Assuming
$\alpha_V(Q^*) \simeq 0.4$, the QCD LO prediction appears to be
smaller by approximately a factor of 2 compared to the presently available data
extracted
from the original pion electroproduction experiments from
CEA.\cite{Bebek:1976ww} A
definitive comparison will require a careful extrapolation to the pion pole and
extraction of the longitudinally polarized photon contribution of the $e p
\to \pi^+ n$ data.

A debate has continued on whether processes such as the pion
and proton
form factors and elastic Compton scattering $\gamma p \to \gamma p$ might be
dominated by higher twist mechanisms until very large momentum
transfers.\cite{Isgur:1989iw,Radyushkin:1998rt,Bolz:1996sw} For example, if one
assumes that the light-cone wavefunction of the pion has the form
$\psi_{\rm soft}(x,k_\perp) = A \exp (-b {k_\perp^2\over x(1-x)})$, then the
Feynman endpoint contribution to the overlap integral at small $k_\perp$ and
$x \simeq 1$ will dominate the form factor compared to the hard-scattering
contribution until
very large $Q^2$.  However, the above form of $\psi_{\rm soft}(x,k_\perp)$
has no
suppression at $k_\perp =0$ for any $x$; \ie, the
wavefunction in the hadron rest frame does not fall-off at all for $k_\perp
= 0$ and
$k_z \to - \infty$.  Thus such wavefunctions do not represent well
soft QCD contributions.  Furthermore, such endpoint contributions will be
suppressed
by the QCD Sudakov form factor, reflecting the fact that a near-on-shell
quark must
radiate
if it absorbs large momentum.  If the endpoint contribution dominates
proton Compton
scattering, then both photons will interact on the same
quark line in a local fashion and the
amplitude is real, in strong contrast to the QCD predictions which have a
complex
phase structure.  The perturbative QCD predictions\cite{Kronfeld:1991kp} for the
Compton amplitude phase can be tested in
virtual Compton scattering by interference with Bethe-Heitler
processes.\cite{Brodsky:1972vv} It should be
noted that there is no apparent endpoint contribution which
could explain the success of dimensional counting in large angle pion
photoproduction.

It is interesting to compare the corresponding calculations of form
factors of bound states in QED.  The soft wavefunction
is the Schr\"odinger-Coulomb solution $\psi_{1s}(\vec k) \propto (1 + {{\vec
p}^2/(\alpha m_{\rm red})^2})^{-2}$, and the full wavefunction,  which
incorporates transversely polarized photon exchange, only differs by
a factor $(1 + {\vec p}^2/m^2_{\rm red})$.  Thus the leading twist dominance of form
factors in QED occurs at relativistic scales $Q^2 > {m^2_{\rm
red}} $.\cite{Brodsky:1989pv}
Furthermore, there are no extra relative factors of $\alpha$ in the
hard-scattering contribution.  If the QCD coupling $\alpha_V$ has an infrared
fixed point, then the fall-off of the valence wavefunctions of hadrons will have
analogous power-law
forms, consistent with the Abelian correspondence
principle.\cite{Brodsky:1997jk} If such power-law wavefunctions are
indeed applicable to the soft domain of QCD then, the transition to
leading-twist
power law behavior will occur in the nominal hard perturbative QCD domain where
$Q^2 \gg \VEV{k^2_\perp}, m_q^2$.

\section*{Acknowledgments}

I thank Carl Carlson, Anatoly Radyushkin, and their colleagues at
Jefferson Laboratory for their kind hospitality at this conference.

\end{document}